\documentclass[twocolumn]{aastex631}
\usepackage{comment, todonotes}
\usepackage{lineno}
\usepackage{tablefootnote}

\newcommand{\Msun}{\ifmmode {M_{\odot}}\else${M_{\odot}}$\fi}
\newcommand{\Rsun}{\ifmmode {R_{\odot}}\else${R_{\odot}}$\fi}
\newcommand{\Lsun}{\ifmmode {L_{\odot}}\else${L_{\odot}}$\fi}
\newcommand{\Rearth}{\ifmmode {R_{\oplus}}\else${R_{\oplus}}$\fi}
\newcommand{\lhal}{\ifmmode {L_{H\alpha}}\else$L_\mathrm{H\alpha}$\fi}

\newcommand{\lapprox}{{\lower0.8ex\hbox{$\buildrel <\over\sim$}}}
\newcommand{\gapprox}{{\lower0.8ex\hbox{$\buildrel >\over\sim$}}}

\def\asec{\ifmmode^{\prime\prime}\else$^{\prime\prime}$\fi}

\newcommand{\gaia}{{Gaia}}
\newcommand{\ispec}{{\tt iSpec}}

\newcommand{\kps}{{{\rm km\ s}^{-1}}}

\newcommand{\Porb}{\ifmmode {P_{\rm orb}}\else${P_{\rm orb}}$\fi}
\newcommand\vsini{\ensuremath{v_{e}\sin i}}
\newcommand\met{$\mathrm{[M/H]}$}
\newcommand{\teff}{${T_{\rm{eff}}}$}
\DeclareRobustCommand{\logg}{$\log g$}

\DeclareRobustCommand{\vmic}{$v_\mathrm{mic}$}
\DeclareRobustCommand{\vmac}{$v_\mathrm{mac}$}
\DeclareRobustCommand{\kms}{$\mathrm{km\, s}^{-1}$}

\newcommand{\Mtot}{\ifmmode {{M_{\rm tot}}}\else{$M_{\rm tot}$}\fi}
\newcommand{\RV}{\ifmmode {{\rm RV}}\else RV \fi}
\newcommand{\bigG}{\ifmmode {\mathcal{G}}\else${\mathcal{G}}$\fi}

\begin{document}
\shorttitle{Chemically peculiar stars in Stock 2}
\shortauthors{Casamiquela et al.}

\title{\sc Chemically Peculiar Stars in the Open Cluster Stock 2}

\newcommand{\columbia}{Department of Astronomy, Columbia University, 550 West 120th Street, New York, NY 10027, USA}
\newcommand{\bordeaux}{Laboratoire d’Astrophysique de Bordeaux, Univ. Bordeaux, CNRS, B18N, all\'ee Geoffroy Saint-Hilaire, 33615 Pessac, France}
\newcommand{\paris}{GEPI, Observatoire de Paris, PSL Research University, CNRS, Sorbonne Paris Cité, 5 place Jules Janssen, 92190 Meudon, France}

\author[0000-0001-5238-8674]{Laia Casamiquela}
\affiliation{\paris}

\author[0000-0002-8675-4000]{Marwan Gebran}
\affiliation{Department of Chemistry and Physics, Saint Mary’s College, Notre Dame, IN 46556, USA}

\author[0000-0001-7077-3664]{Marcel A.~Ag\"{u}eros}
\affiliation{\columbia}

\author[0000-0002-7084-487X]{Herv\'e Bouy}
\affiliation{\bordeaux}

\author[0000-0003-3304-8134]{Caroline Soubiran}
\affiliation{\bordeaux}

\begin{abstract}
The recently re-discovered open cluster Stock 2, located roughly 375 pc away and about 400 Myr old, has the potential to be an exciting new testbed for our understanding of stellar evolution. We present results from a spectroscopic campaign to characterize stars near the cluster's main-sequence turnoff; our goal is to identify candidate chemically peculiar stars among the cluster's A stars. We obtained \'echelle spectra for 64 cluster members with ESPaDOnS on the 3.6-m Canada-France-Hawaii Telescope, Mauna Kea Observatory, USA, and for six stars with SOPHIE on the 1.93-m telescope at the Observatoire de Haute-Provence, France. We complemented these new observations with those of 13 high-mass cluster members from the HARPS-N archive; our overall sample is of 71 stars. We derived the fundamental parameters (\teff, \logg, \met) as well as \vsini\ for our sample using the Sliced Inverse Regression (SIR) technique, and then used {\tt iSpec} to derive individual abundances of 12 chemical species. With these abundance determinations, we identified nine A stars with anomalous levels of Sc, Ca, and other metallic lines. Follow-up observations of these Am candidates with a known age can transform them into benchmarks for evolutionary models that include atomic diffusion and help build a better understanding of the complex interactions between macroscopic and microscopic processes in stellar interiors.
\end{abstract}


\section{Introduction}
\label{sec:intro}

Spectroscopic studies of intermediate-mass Main Sequence (MS) stars ($\gapprox$1.2~\Msun, or spectral types B, A, and F) have found that some fraction are chemically peculiar, at least in their upper atmospheric layers \citep[e.g.,][]{1974ARA&A..12..257P}. A subset of these stars are known as metallic-line stars, or Am stars, and are characterized by an underabundance of light elements such as C, O, Ca, Sc, and an overabundance of iron and iron-peak/neutron-capture elements such as Sr, Y, and Ba \citep{coma,pleiades,Hyades,monier18}. 
Am stars are usually slow rotators in comparison to normal A stars, with a projected equatorial rotational velocity $\vsini \lesssim100$~\kms\ \citep{1974ARA&A..12..257P,royer14}, and a significant fraction are found in binary systems \citep[e.g.,][]{abt1973}. Finally, these stars typically have very weak to undetectable magnetic fields, further distinguishing them from Ap stars, which have magnetic fields that can reach values of several kG, and do not generally have companions \citep{2007A&A...475.1053A,2007AstBu..62...62R,2018CoSka..48...48B,2020MNRAS.492.5794B}.

The chemical peculiarities that characterize Am stars are often explained by invoking atomic diffusion models \citep{Michaud1970}. Correctly accounting for the competition between radiative levitation and gravitational settling, and for the impact of other rotational instabilities, such as turbulent transport and mass loss, can produce their peculiar abundance patterns  \citep{2000IAUJD...5E...7M,2000ApJ...529..338R,2001ApJ...558..377R,2005ApJ...623..442M,Talon2006,vick2010,2011A&A...526A..37V,deal20}. Other mechanisms that have been proposed to explain the formation of chemically peculiar stars include pollution due to mass transfer from a binary companion \citep{vandenHeuvel1968}, tidal interactions with hot brown dwarfs or hot Jupiters \citep[][]{Saffe+2022}, and planet engulfment \cite{Church+2020}.

To understand the formation of chemically peculiar stars, and constrain evolutionary models, accurate abundance measurements for Am stars of known ages are indispensable \citep{1974ARA&A..12..257P,2020ApJ...898...28X,deal20}. High-resolution spectroscopy of Am stars in open clusters is therefore crucial for advancing our understanding of the processes that produce these abundance patterns.

\begin{figure*}[!t]
\centerline{\includegraphics[trim=.2cm 0cm .5cm .5cm, clip=True, scale=0.65]{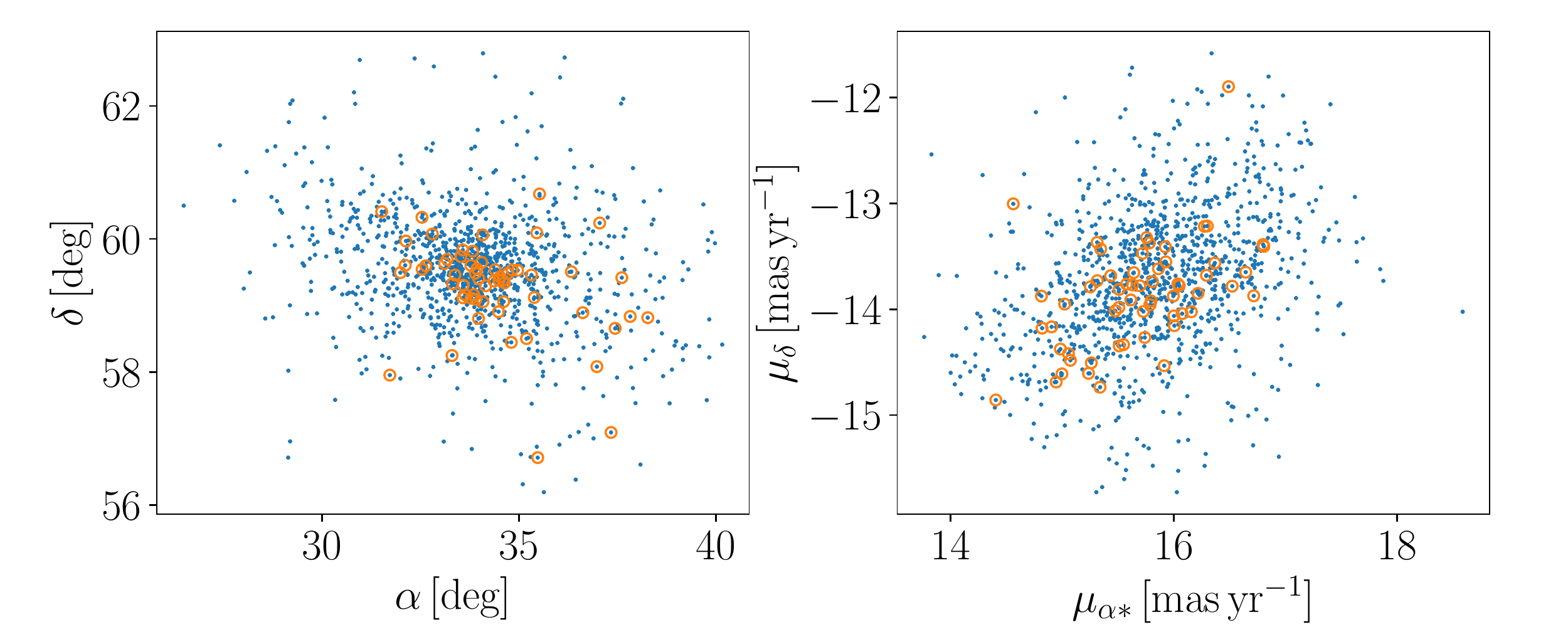}}
\caption{{\it Left---}Spatial distribution of Stock 2 members (blue), with our spectroscopic targets highlighted (orange). {\it Right---}Proper motion distribution.}
\label{fig:gdr3}
\end{figure*}

Stock 2 \citep{1956ApJ...123..258S}, an open cluster in the Orion spiral arm, has not been extensively studied, despite being relatively nearby. \cite{spagna2009} estimated the distance to the cluster to be $\approx$350 pc, its age to be between 200 and 500~Myr, and the average reddening for the cluster to be $E(B-V)=0.30$. Using data from the first two Gaia 
data releases \citep[DR1, DR2;][]{gaia16,gaia18}, \cite{reddy19} derived a distance of 372$-$390 pc, an age of 225 Myr, an average reddening of $E(B-V)=0.45$, and an overall metallicity [Fe/H]~$=-0.06$~dex.
\citet{Cantat-Gaudin+2020} applied an artificial neural network to \gaia\ DR2 photometry and astrometry to estimate distances, ages, and reddenings for a large sample of known clusters. For Stock~2, these authors obtained a list of 1200 members down to $G=18$ mag, and determined an age of 400~Myr, heliocentric distance of 435~pc, and an average extinction $A_V=0.5$.

Two recent studies have provided additional measurements of Stock 2's metallicity. \cite{2021AJ....161....8Y} found [Fe/H]~$=-0.040\pm 0.147$~dex for a sample of cluster members in the  medium-resolution LAMOST data release~7 catalog. \cite{2021A&A...656A.149A} derived the stellar parameters, extinction, and radial and projected rotational velocities for 46 dwarf and giant star members of the cluster. These authors estimated its age to be 450$\pm$150~Myr, and its average  [Fe/H]~$=0.07$$\pm$$0.06$~dex. \cite{2021A&A...656A.149A} also concluded that the extended MS turnoff in this cluster (see the cluster color-magnitude diagram [CMD], Figure~\ref{fig:cmd}) is most probably due to differential reddening, with the average value being $E(B-V)=0.27$$\pm$$0.11$.

Our aim is to chemically characterize the hot MS A stars in this cluster, and thereby investigate the presence of chemically peculiar stars, particularly of Am stars. 
This paper is structured as follows: in Section~\ref{sec:cmd}, we describe our target selection and we analyse the differential reddening in the region of the cluster. In Section~\ref{sec:obs}, we describe our observations of turn-off stars in Stock 2, as well as our use of archival spectroscopic observations of cluster members. In Section~\ref{sec:AP}, we derive the atmospheric parameters for the stars in our sample. Section~\ref{sec:abund} describes the abundance computation method and its results. In Section~\ref{sec:discussion}, we discuss the presence of chemically peculiar stars among our targets, and we compare our results to those of similar studies of different open clusters. We conclude in  Section~\ref{sec:conclusion}.

\begin{figure}[!h]
\centerline{\includegraphics[trim=.8cm .2cm 1.8cm .25cm, clip=True, scale=0.46]{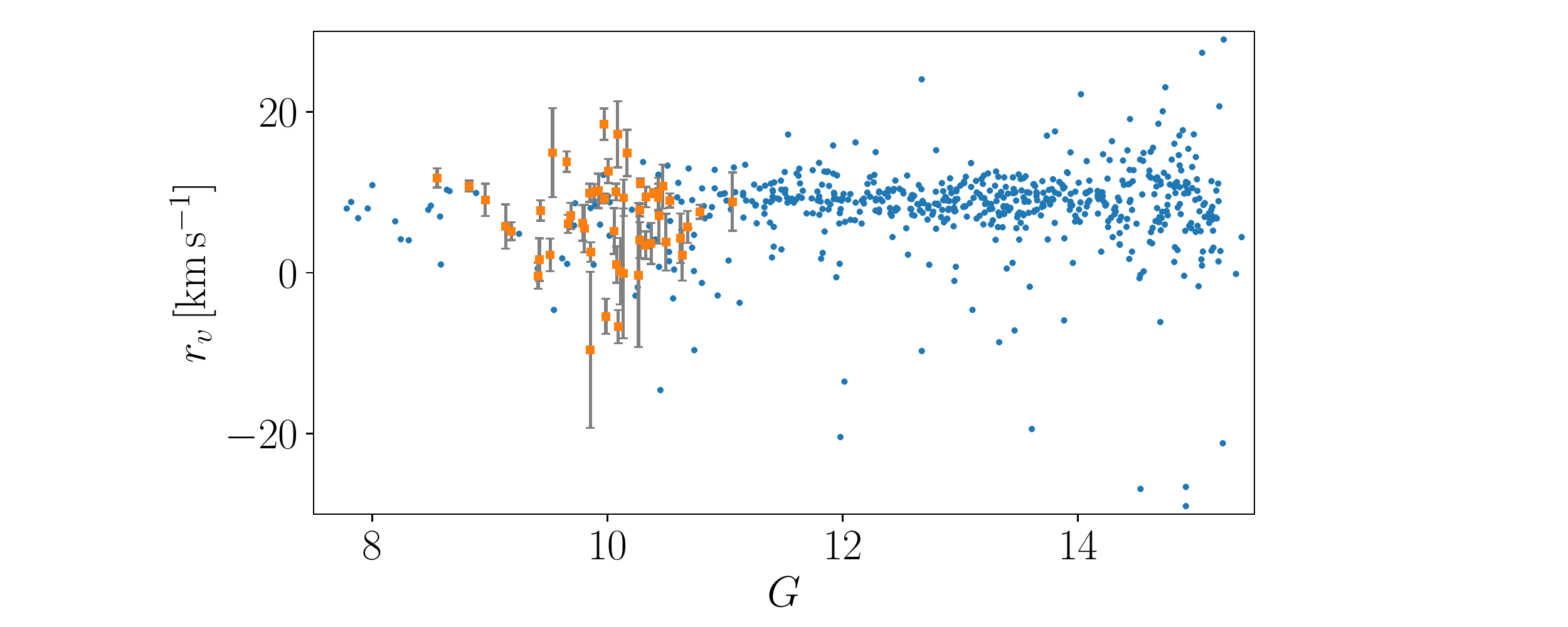}}
\caption{Gaia DR3 RVs for Stock 2 members and for our spectroscopic targets, as a function of $G$ magnitude. Uncertainties for the members are not plotted for clarity, they have a mean value of 3 \kms. While the RVs for our targets are consistent with the median RV for the cluster, the dispersion in the values is larger than for fainter members, as is expected for these hotter stars.}
\label{fig:rvs}
\end{figure}

\section{Target selection and color-magnitude diagram} \label{sec:cmd}
We relied on the Stock 2 membership list constructed by \citet{Cantat-Gaudin+2020}, and we crossmatched it with Gaia DR3 data \citep[][]{GaiaDR3}. We plot in Figure~\ref{fig:gdr3} the spatial and proper motion distributions of the member stars and highlight our spectroscopic targets. Our targets were selected to be among the highest probability members on the cluster's upper MS, as determined by \citet{Cantat-Gaudin+2020} from Gaia DR2 astrometry.

\begin{figure*}[!th]
\centerline{\includegraphics[trim=.2cm .2cm .5cm .5cm, clip=True, scale=0.65]{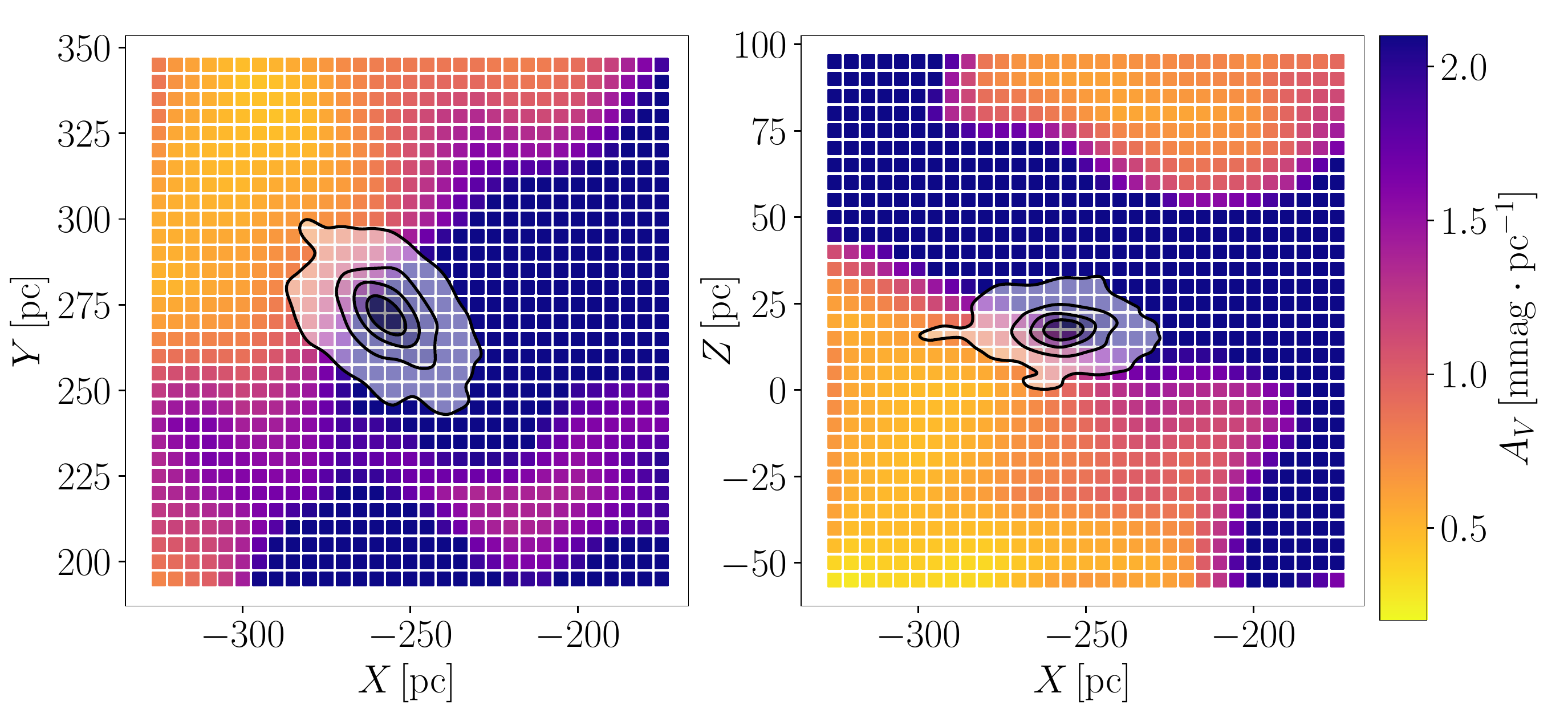}}
\caption{Extinction map used to produce a dereddened CMD for Stock 2. The cluster members' positions in Cartesian space (centered on the Sun) are indicated by the contours. Only one layer of the data cube is shown in each projection. }
\label{fig:extinction}
\end{figure*}

In Figure~\ref{fig:rvs}, we plot the Gaia DR3 radial velocity (RV) distribution of the member stars and of our targets as a function of $G$ magnitude. The median RV of the cluster and its median absolute deviation is $8.8$$\pm$$3.8\,\kps$. The RVs of the spectroscopic targets are centered on this value, but have a larger dispersion 
compared to those of the fainter members. As explained by \citet{Bloome+2022}, the Gaia spectra of hot stars are dominated by strong Ca and H Paschen lines, which are not optimal for RV determinations. This results in larger RV uncertainties and introduces magnitude-dependent systematic effects. We corrected for the systematics using the formula in \citet{Bloome+2022}, but the large dispersion in RVs for these hotter stars is still very obvious in Figure~\ref{fig:rvs}\footnote{Binarity can also increase the RV dispersion.}. For the mentioned reasons we do not consider these RV to settle the membership of the spectroscopic targets.

In addition, as pointed out by \cite{2021A&A...656A.149A}, Stock 2 is located in a region affected by differential reddening. To produce a dereddened CMD of the cluster, we used the 3D extinction map of \citet{Lallement+2022}. These authors combined EDR3 with 2MASS data \citep{twomass} to derive the extinction toward stars with accurate photometry and relative EDR3 parallax uncertainties of less than 20\%. The resulting map covers a volume of 6$\times$6$\times$0.8~kpc$^3$ centered on the Sun with a resolution of 5 pc. This is sufficient for our purposes, as Stock 2 has an extent of about $\approx$80~pc. 

We computed the location of each star in Cartesian $(X,Y,Z)$ coordinates using the direct inversion of the DR3 parallax. We then linearly interpolated the extinction map to find the value of $A_V$ integrated along the line of sight. We corrected the three Gaia DR3 magnitudes using the $\frac{A_{\lambda}}{A_V}$ coefficients listed on the PARSEC website.\footnote{\url{http://stev.oapd.inaf.it/cgi-bin/cmd_3.7}} We plot in Figure~\ref{fig:extinction} the resulting extinction map for a region centered on Stock 2.

\begin{figure}[h]
\centerline{\includegraphics[trim=.2cm .2cm .75cm .95cm, clip=True, scale=0.52]{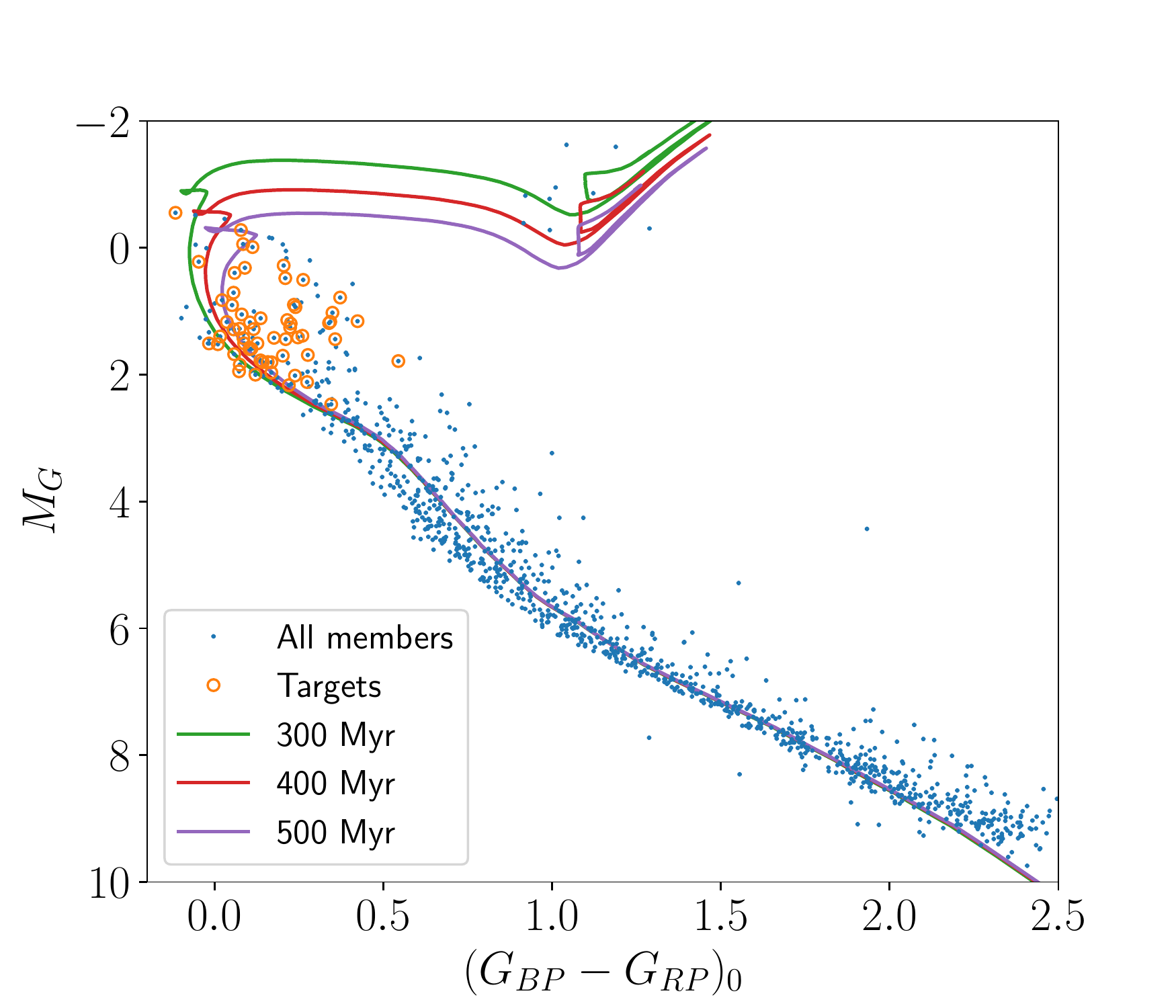}}
\caption{Absolute and derreddened \gaia\ DR3 CMD for Stock 2, with the members derived by  \cite{Cantat-Gaudin+2020} in blue and our spectroscopic targets indicated with orange circles. We also show a three isochrones of different ages and Solar metallicity for reference.}
\label{fig:cmd}
\end{figure}

The absorption values for individual stars vary between 0.4 and 1.2 mag.
We plot in Figure~\ref{fig:cmd} the dereddened CMD of the cluster. We overplot three Solar metallicity PARSECv2.0 isochrones \citep[][]{Bressan+2012,Nguyen+2022} corresponding to ages of 300, 400, and 500 Myr, roughly the range of ages quoted for Stock~2 in the literature.

\begin{figure*}[!thp]
\centerline{\includegraphics[trim=4.5cm 2.75cm 5cm 4.5cm, clip=True, scale=0.4]{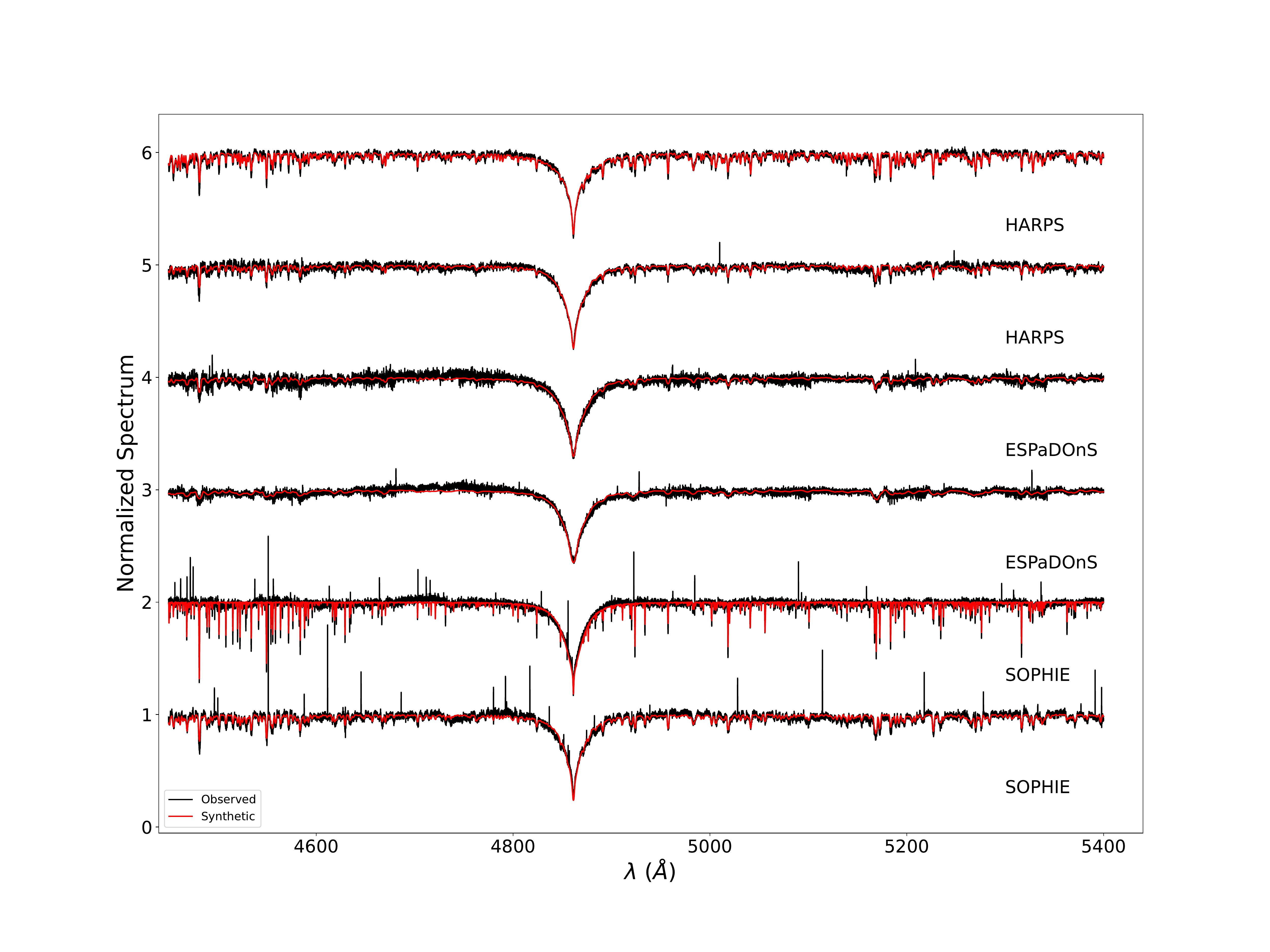}}
\caption{Example spectra for stars in our sample obtained with the SOPHIE, ESPaDOnS, and HARPS spectrographs. The wavelength range, between 4450 and 5400 \AA, corresponds to the one used to derive stellar parameters.  The spectra (in black) are shifted vertically for clarity; we also show the best-fit synthetic spectra (red) for each.}
\label{fig:best-fit}
\end{figure*}

As mentioned in Section~\ref{sec:intro}, the CMD of the cluster presents an extended MS turnoff that cannot be  explained simply by the impact of stellar rotation \citep{2021A&A...656A.149A}. Our extinction-corrected CMD does show a thinner MS with respect to the non-corrected CMD. The isochrone match (by eye) to both the MS and the red clump is also much better. 
However, we still obtained a noticeably extended MS turnoff, which indicates that this feature cannot be fully explained by the differential extinction captured by our map.

\section{Observations and data reduction} \label{sec:obs}

We used a combination of our own observations and archival spectra to conduct our search for chemically peculiar stars in Stock~2. Our targets' properties are listed in Table~\ref{tab:results}, and sample spectra are shown in Figure~\ref{fig:best-fit}. We describe below the different instruments and reductions pipelines used in our study.

\subsection{ESPaDOnS}\label{sec:espadon}
During the 2020B semester, we observed 64 hot Stock~2 MS stars with the ESPaDOnS spectrograph mounted on the 3.6-m Canada-France-Hawaii telescope (CFHT) located at the Mauna Kea Observatory, USA (PI: Casamiquela, ID: 20BF003). We used the Star+Sky mode, which gives a full optical spectrum (3700--10,500~\AA) with a mean resolution $R = 68,000$. The data were reduced using the instrument pipeline (Upena version 1.1). 

\subsection{SOPHIE}\label{sec:sophie}
We complemented these observations with those of six MS stars re-observed on the night of 2022 Feb 8 using the SOPHIE spectrograph, a cross-dispersed \'echelle spectrograph mounted on the 1.93-m telescope of the Observatoire de Haute Provence, France (PI: Bouy, ID: 21B.PNPS.BOUY). We used the High Efficiency mode, leading to spectra over the wavelength range 3870--6940~\AA\ in 39 orders with $R = 40,000$. The data were processed automatically by the instrument pipeline \citep{Perruchot+2008}.

\subsection{HARPS}
Finally, we analyzed the spectra of 13 MS stars observed by \cite{2021A&A...656A.149A} using the HARPS-N \'echelle spectrograph mounted on the 3.6-m Telescopio Nazionale Galileo at El Roque de los Muchachos Observatory, Spain. These spectra cover the wavelength range $\approx$4000--7000 \AA\ with $R = 115,000$. The data reduction was performed using the instrument pipeline.

\section{Atmospheric Parameters}\label{sec:AP}
We derived effective temperature (\teff), surface gravity (\logg), metallicity (\met), and \vsini\ using a combination of Principal Component Analysis (PCA, \citealt{Gebran16}) and Sliced Inverse Regression (SIR, \citealt{Kassounian19}) techniques. In brief, the method is based on a dimensionality reduction tool that makes the comparison of the observed spectrum with a trained synthetic database of spectra.
Details about our method are provided in Sections~\ref{sec:TD} and \ref{sec:SIR}, and our results are described in Section~\ref{sec:results}.

\begin{table}[h]
\centering
\begin{tabular}{|c|c|c|}
    \hline
    Parameter& Range & Step  \\ \hline
    \teff & $5000 \to 11,000$ K & 50 K \\
    \logg & $2.0 \to 5.0$ dex & 0.05 dex \\
    \met & $-1.5 \to +1.5$ dex & 0.01 dex \\
    \vsini & $0 \to 300$ \kms & 1 \kms \\
    $\lambda$ & $4450 \to 5500$ \AA & $\lambda/\mathrm{Resolution}$ \\ \hline
\end{tabular}
\caption{Range of the stellar parameters used in the construction of the training databases.}\label{tab:stellar-range}
\end{table}

\subsection{Training database}\label{sec:TD}
We used three databases, with different resolving power, in this work. Each database consists of a set of $\sim$80,000 synthetic spectra with stellar parameters randomly selected in the ranges given in Table~\ref{tab:stellar-range}. The wavelength range, between 4450 and 5400 \AA, was selected because it is very sensitive to all the stellar parameters, especially for A, F, and G stars \citep{S4n,Dm,Gebran16,Kassounian19}. This is mainly due to the presence of the Balmer H$\beta$ line and of moderately strong and weak metallic lines in this window, and is helped by the existence of high precision atomic data for the available transitions.\footnote{Non-LTE effects are negligible in this region due to the absence of strong metallic lines \citep{2021osvm.confE...5M,2020MNRAS.499.3706M,2020MNRAS.493.6095M,2018ApJ...866..153A}.} 
The microturbulent velocity parameter $\xi_t$ is selected between 2 and 4 \kms, which are typical values for A and Am stars  \citep{micro}. The studied region is only moderately affected by variations in this parameter.

We calculated atmospheres using the 1D plane-parallel ATLAS9 models \citep{Kurucz1992}. These models account for local thermodynamic equilibrium (LTE) layers in hydrostatic and radiative equilibrium. We implemented the new opacity distribution function of \citet{castelli}, and the mixing length parameter was used as prescribed by \cite{2004IAUS..224..131S}. Synthetic spectra were calculated using the SYNSPEC48 code \citep{spectra}. We used the line list compiled in \cite{CNN} based on the three databases gfhyperall\footnote{\url{http://kurucz.harvard.edu}}, VALD\footnote{\url{http://vald.astro.uu.se}}, and NIST\footnote{\url{http://physics.nist.gov}}.

\subsection{Sliced Inverse Regression}\label{sec:SIR}
SIR requires the pre-processing of the training database using PCA. A detailed description of the PCA technique and its calculation steps can be found in \cite{Gebran16}. In summary, the training database was gathered into a matrix of dimension $N_{\mathrm{Spectra}} \times N_{\mathrm{\lambda}}$, where $N_{\mathrm{Spectra}}$ was the number of spectra in each database ($\sim$80,000) and $N_{\mathrm{\lambda}}$ was the number of wavelength points per synthetic spectrum.
The covariance matrix $\mathrm{\textbf{C}}$ was then calculated as follows:

\begin{equation}
    \mathrm{\textbf{C}}=\frac{1}{N_{\mathrm{Spectra}}} \sum_{i=1}^{N_{\mathrm{Spectra}}} (x_i - \bar{x})\cdot (x_i - \bar{x})^T.
\label{eq:cov}
\end{equation}

The superscript $T$ stands for the transpose operator, and $\bar{x}$ is the average spectrum calculated using the classical function
\begin{equation}
    \bar{x}=\frac{1}{N_{\mathrm{Spectra}}} \sum_{i=1}^{N_{\mathrm{Spectra}}} x_i.
    \label{eq:average-spectrum}
\end{equation}

The covariance matrix $\mathrm{\textbf{C}}$ has a dimension of $N_{\mathrm{\lambda}} \times N_{\mathrm{\lambda}}$. The $N_{\mathrm{\lambda}}$ eigenvectors of $\mathrm{\textbf{C}}$, $\mathrm{e_k}$, were then derived and only the first 12 vectors were considered in our study. The choice of 12 eigenvectors was based on the analysis of the reconstructed error: choosing $k$~=~12 leads to a reconstructed error of less than 1\% \citep{Gebran16,Kassounian19}. The synthetic spectra of the database were then projected on these 12 eigenvectors and 12 coefficients ($p_k$) were derived for each spectrum. 

In the same way, observations were projected onto $\mathrm{e_k}$ and the 12 coefficients ($\rho_k$) were derived for each observation. Finally, the values of the stellar parameters of the observation were the ones of the nearest neighbor, found by minimizing the difference:
\begin{equation}
     d_j=\sum_{k=1}^{12} (\rho_k - p_{jk})^2,
\label{eq:minim}
 \end{equation}
where $j$ covers the number of synthetic spectra. 

\begin{table*}[htp]
\centering
\caption{Atmospheric parameters and abundances of the analyzed spectra (the full table is available online). The spectra are identified by \gaia\ DR2 source ID of the star, and we indicate the relevant spectrograph (SOPHIE, HARPS or ESPaDOns) and the spectrum's SNR. The  atmospheric parameters (\teff, \logg, \vmic) and \vsini\ are those computed by SIR. Abundances  computed by \ispec\ with respect to the Sun are also included, together with their uncertainties.}\label{tab:results}
\setlength\tabcolsep{1.5pt}
\begin{tabular}{lccccccccccccc}
\hline
Source ID & Instr. & SNR & \teff  & \logg  & \vsini & [Fe/H] & [Ti/H] & [Ca/H] & [Sc/H] & [Ba/H] & ... \\
 &  &  &  (K) &  & (\kms) & (\kms) &  &  &  &  &  &  \\
\hline
\hline
459223645670638080 & SOPHIE  & 89 & 9290 & 3.64 & 54.0 & 0.03$\pm$0.08 & $-0.23$$\pm$$0.12$ & $-0.24$$\pm$$0.33$ & $-0.12$$\pm$$0.07$ & $0.62$$\pm$$0.20$ & ... \\
458857954974047360 & ESPADONS& 55 & 8073 & 4.29 & 52.3 & 0.02$\pm$0.18 & $-0.18$$\pm$$0.14$ & $-0.36$$\pm$$0.31$ & $-0.27$$\pm$$0.07$ & $0.30$$\pm$$0.33$ & ... \\
459037931282246912 & ESPADONS& 42 & 8274 & 4.61 & 34.2 & 0.25$\pm$0.11 &  0.32$\pm$0.11 & 0.57$\pm$0.10 &  0.57$\pm$0.10 & 0.30$\pm$0.07 & ... \\
459105306430577408 & SOPHIE  & 88 & 8101 & 3.68 & 77.2 & 0.01$\pm$0.13 & $-0.04$$\pm$0.11 & $-0.03$$\pm$0.20 & $-0.26$$\pm$0.07 & 0.31$\pm$0.26 & ... \\
459067441998085376 & ESPADONS& 40 & 8871 & 3.97 &  6.8 & $-0.08$$\pm$0.05 & $-0.05$$\pm$0.10 & $-0.17$$\pm$0.01 & $-0.21$$\pm$0.02 & 0.68$\pm$0.13 & ... \\
507116585468397056 & HARPS   & 83 & 9105 & 4.57 & 33.4 & 0.07$\pm$0.20 &  0.14$\pm$0.22 & 0.15$\pm$0.12 &  0.22$\pm$0.07 & 0.71$\pm$0.12 & ... \\
507222619614549120 & HARPS   & 63 & 7574 & 4.69 & 59.9 & $-0.22$$\pm$0.20 &  0.12$\pm$0.28 & -                &  0.15$\pm$0.29 & -               & ... \\

\hline
\end{tabular}
\end{table*}

During the minimization procedure, the continuum of the observed spectrum was corrected iteratively by applying the \cite{Gazzano10} procedure, as detailed in \cite{Gebran16}. These stellar parameters were used as a starting point for SIR.

SIR starts by calculating the covariance matrix as done in Equation~\ref{eq:cov}. As stellar parameters are derived one by one, we sorted the synthetic spectra by increasing order of the parameter we wished to derive. As an example, if we were deriving \teff, we sorted the synthetic spectra matrix in \teff\ while keeping \logg, \met, and \vsini\ randomly distributed. 

Slices were built by combining subsets of these spectra having similar or close values for the considered parameter. We then calculated the average spectrum, $\bar{x}_h$, in each slice:
\begin{equation}
    \bar{x}_h = \frac{1}{n_h}\sum_{x \in S_h} x_i,
\end{equation}
where $S_h$ is the slice having the index $h$ and $n_h$ is the number of synthetic spectra in that slice.

The next step was to calculate the intra-slice covariance matrix $\Gamma$:

\begin{equation}
\Gamma = \sum_{h=1}^{H} \frac{n_h}{N_{\mathrm{spectra}}} (\bar{x}_h - \bar{x})\cdot (\bar{x}_h - \bar{x})^T,    
\end{equation}
where $H$ is the total number of slices and $\bar{x}$ is the average spectrum defined in Equation~\ref{eq:average-spectrum}.

The final step was to calculate the matrix $\mathrm{\textbf{C}}\Gamma^{-1}$ and evaluate its eigenvector corresponding to the largest eigenvalue. This eigenvector was used for the inversion procedure and for the derivation of parameters (equation~9 of \citealt{Kassounian19}). The role of the PCA in this technique was to reduce the dimension of the initial training database by selecting the first few $\sim$1000 spectra that are the closest, in terms of $d$ (Equation~\ref{eq:minim}), to the observed one. This new matrix was then used to calculate the covariance matrix used, in turn, by SIR. 

\subsection{Resulting stellar parameters}\label{sec:results}
The stellar parameters derived using SIR are listed in Table~\ref{tab:results}. The average errors are 200 K, 0.20~dex, 0.15~dex, and 2~\kms\ for \teff, \logg, \met, and \vsini, respectively. The metallicty is just an estimation of the overall metal content in the spectral region analyzed by SIR. The line-by-line derivation of the metal content is done in Section~\ref{sec:abund}, where we have used well-studied lines with accurate atomic data. The signal-to-noise ratio (SNR) is derived for all spectra between 4450 and 5500~\AA\ using \cite{snr_der}'s \texttt{DER\_SNR}\footnote{The name is taken from the FITS header keyword name used for the “derived SNR” in the VO Spectral Data Model (\url{http://www.ivoa.net/Documents/latest/SpectrumDM.html}).} algorithm. In Figure~\ref{fig:best-fit}, we show examples of the best fit synthetic spectra with SIR-derived parameters plotted with spectra obtained with the three different spectrographs.

\begin{center}
    \begin{figure}[!t]
        \centerline{\includegraphics[trim=1.5cm 2.75cm 3cm 3.5cm, clip=True, width=1.05\columnwidth]{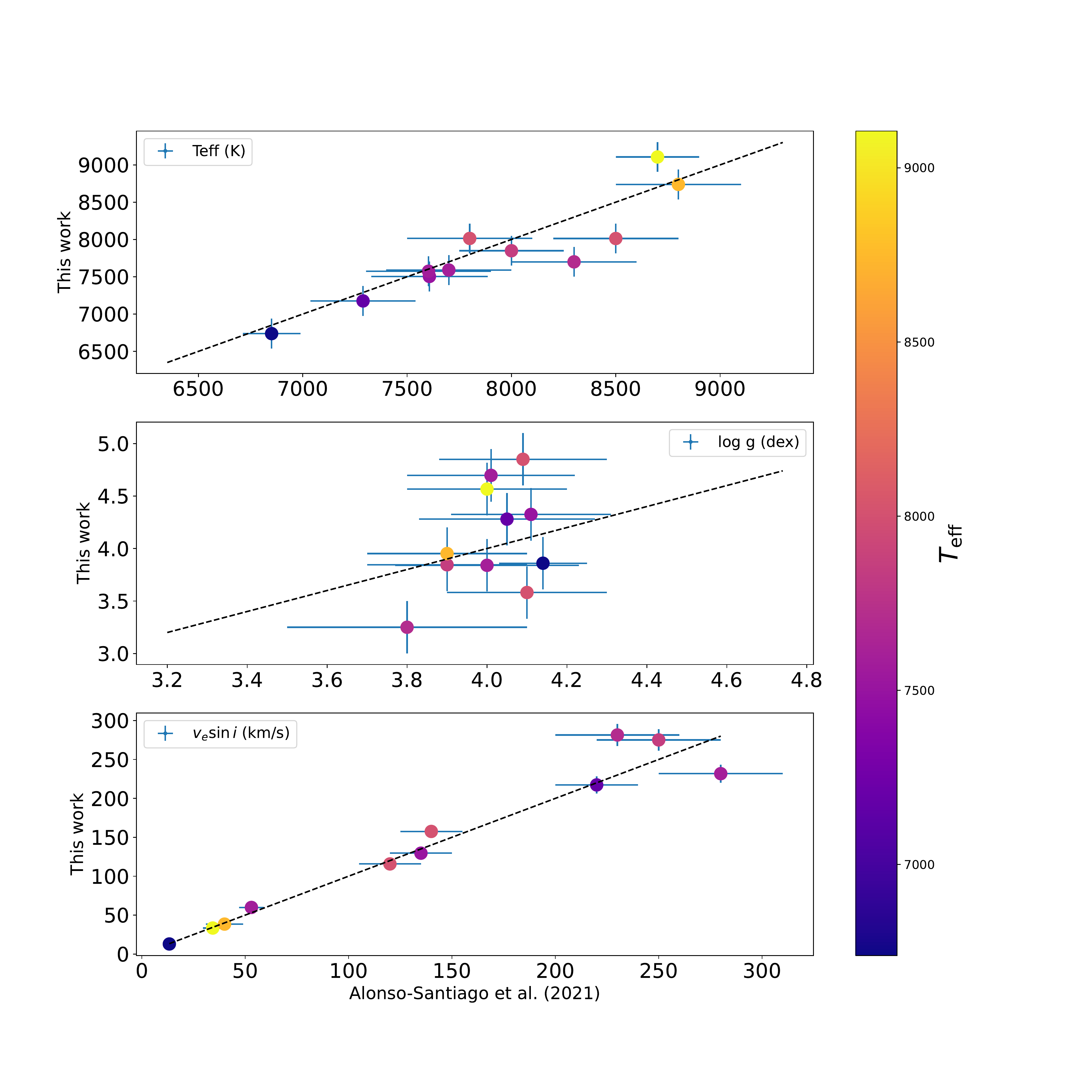}}
        \caption{Comparison between the \teff, \logg, and \vsini\ values found in this work and the ones derived by \cite{2021A&A...656A.149A} for the common 11 stars. The agreement in \teff\ and \vsini\ is very good; that in \logg\ is not, but this is unsurprising, as different methods for deriving spectroscopic \logg\ values for A stars are known to produce quite different values. }
        \label{fig:comp}
    \end{figure}
\end{center}

We also compared our results to those of \cite{2021A&A...656A.149A} for the 11 A stars our samples have in common. The results are shown in Figure~\ref{fig:comp} for \teff, \logg, and \vsini. The agreement is very good for \teff\ and \vsini\ (top and bottom panels), but we found larger disagreements for \logg\ (middle panel). This is unsurprising, as uncertainties on \logg\ derived from spectroscopy are known to be of order 0.15-0.3 dex for A stars \citep{2005MSAIS...8..130S}, and one should therefore expect significant differences when comparing \logg\ values derived using different techniques. For consistency, we used the values we obtained using SIR for the derivation of the individual chemical abundances in Section~\ref{sec:abund}.

We observed nine stars with more than one instrument: eight with SOPHIE and ESPaDOns, and one with all three spectrographs. We derived stellar parameters for each spectrum, and include the results in Table~\ref{tab:results}. Our \teff\ and \vsini\ values for these multiply observed stars agree within the uncertainties: the average difference between the derived \teff\ is around 150 K and between the \vsini, 7~\kms. The variation in \logg\ is significant for some stars, especially Gaia DR2 459105306430577408 (1.2~dex) and Gaia DR2 507514058918803328 (0.6 dex). This is probably due to the sensitivity of the \logg\ measurement to the SNR; in these cases the difference in SNR between the spectra is large.

\section{Abundances}\label{sec:abund}
Using the atmospheric parameters obtained as described in the previous section, we derived 1D-LTE chemical abundances of 12 chemical species using \ispec\ \citep{BlancoCuaresma+2014,BlancoCuaresma2019}: O, Mg, Si, Ca, Sc, Ti, Cr, Fe, Sr, Y, Zr, and Ba. We then computed abundances for a subsample of 20 stars where spectral lines were measurable. In the other cases, the SNR was too low, and/or the star's rotational velocity was too high, and we could not identify the lines to perform the synthetic fit. Below we describe our method (Section~\ref{sec:ispec}) and our results (Section~\ref{sec:results_abund}).

\begin{table*}[t]
\centering
\caption{Selection of clean lines used for the abundance computation. For each element we list the available lines and include the wavelength peak, the excitation potential, and the logarithm of the oscillator strength taken from \cite{Heiter+2021} and \cite{Gebran16}.}\label{tab:lines}
\begin{tabular}{lccc|lccc|lccc}
\hline
Element & $\lambda_{\mathrm{peak}}$ (nm) & $\chi$ (eV) & $\log gf$  & Element & $\lambda_{\mathrm{peak}}$ (nm) &$\chi$ (eV) & $\log gf$ & Element & $\lambda_{\mathrm{peak}}$ (nm) &$\chi$ (eV) & $\log gf$  \\
\hline
\hline
O I   & 436.8258    & 9.521       &$-$2.186   & Ti II  & 454.9621     & 1.584      & $-$0.220  & Fe II  & 451.5339    & 2.844      &  $-$2.450 \\
Mg II  & 439.0572    & 9.999      &$-$0.530   & Ti II  & 457.1971    & 1.572      & $-$0.310  & Fe II  & 452.0224    & 2.807      &  $-$2.600 \\
Si II  & 504.1024    & 10.067     & 0.150   & Ti II  & 465.7200    & 1.243      & $-$2.290  & Fe II  & 452.2634    & 2.844      &  $-$2.030 \\
Si II  & 505.5984    & 10.074     & 0.530   & Ti II  & 480.5085    & 2.061      & $-$0.960  & Fe II  & 454.1524    & 2.856      &  $-$2.790 \\
Ca II  & 500.1479    & 7.505      &$-$0.506   & Ti II  & 512.9156    & 1.892      & $-$1.340  & Fe II  & 455.5890    & 2.828      &  $-$2.160 \\
Ca II  & 501.9971    & 7.515      &$-$0.247   & Ti II  & 518.8687    & 1.582      & $-$1.050  & Fe II  & 457.6340    & 2.844      &  $-$2.920 \\
Sc II  & 432.0732    & 0.605      & $-$0.252  & Ti II  & 533.6786    & 1.582      & $-$1.600  & Fe II  & 462.0521    & 2.828      &  $-$3.240 \\
Sc II  & 467.0407    & 1.357      & $-$0.576  & Cr II  & 455.8650    & 4.073      &  $-$0.449 & Fe II  & 465.6981    & 2.891      &  $-$3.610 \\
Sc II  & 503.1021    & 1.357      & $-$0.400  & Cr II  & 458.8199    & 4.071      &  $-$0.627 & Fe II  & 466.6758    & 2.828      &  $-$3.368 \\
Sc II  & 523.9813    & 1.455      & $-$0.765  & Cr II  & 459.2049    & 4.074      &  $-$1.221 & Fe II  & 492.3927    & 2.891      &  $-$1.260 \\
Sc II  & 565.7896    & 1.507      & $-$0.603  & Cr II  & 461.6629    & 4.072      &  $-$1.361 & Fe II  & 519.7577    & 10.398     &  $-$2.105 \\
Ti II  & 429.0219    & 1.165      & $-$0.870  & Cr II  & 461.8803    & 4.074      &  $-$0.840 & Fe II  & 531.6615    & 3.153      &  $-$1.870 \\
Ti II  & 430.0042    & 1.180      & $-$0.460  & Cr II  & 463.4070    & 4.072      &  $-$0.990 & Fe II  & 550.6195    & 10.522     &  0.950 \\
Ti II  & 439.4059    & 1.221      & $-$1.770  & Cr II  & 481.2337    & 3.864      &  $-$1.977 & Sr II  & 421.5520    & 0.000      & $-$0.145 \\
Ti II  & 439.5031    & 1.084      & $-$0.540  & Cr II  & 523.7329    & 4.073      &  $-$1.144 & Y II   & 488.3684    & 1.084      & 0.190 \\
Ti II  & 439.9772    & 1.237      & $-$1.200  & Cr II  & 531.3590    & 4.074      &  $-$1.526 & Y II   & 490.0120    & 1.033      & 0.030 \\
Ti II  & 441.1072    & 3.095      & $-$0.650  & Fe II  & 423.3172    & 2.583      &  $-$1.900 & Zr II  & 420.8980    & 0.713      & $-$0.510 \\
Ti II  & 441.7714    & 1.165      & $-$1.190  & Fe II  & 425.8154    & 2.704      &  $-$3.478 & Zr II  & 449.6960    & 0.713      & $-$0.890 \\
Ti II  & 444.3801    & 1.080      & $-$0.710  & Fe II  & 429.6572    & 2.704      &  $-$2.933 & Ba II  & 455.4029    & 0.000      &  0.170 \\
Ti II  & 446.8492    & 1.131      & $-$0.630  & Fe II  & 438.5387    & 2.778      &  $-$2.680 & Ba II  & 493.4076    & 0.000      &  $-$0.157 \\
Ti II  & 448.8325    & 3.124      & $-$0.500  & Fe II  & 449.1405    & 2.856      &  $-$2.700 & Ba II  & 614.1713    & 0.704      &  $-$0.032 \\
Ti II  & 450.1270    & 1.116      & $-$0.770  & Fe II  & 450.8288    & 2.856      &  $-$2.250 & Ba II  & 649.6897    & 0.604      &  $-$0.407 \\
\hline
\end{tabular}

\end{table*}

\subsection{Method}\label{sec:ispec}
We ran \ispec\ through a pipeline adapted from that of \citet{Casamiquela+2020}. We used the synthetic spectral synthesis fitting technique, running the radiative transfer code SPECTRUM \citep{Gray+1994} and the ATLAS9 atmospheric models \citep{castelli}, which are implemented in \ispec. We used both the \gaia-ESO survey linelist \citep{Heiter+2021} and the selection of lines used in \cite{coma}, \cite{pleiades}, \cite{Hyades}, and \cite{monier18}, which are adapted for A and F-type stars and were used in the derivation of the stellar parameters in Section~\ref{sec:AP}. We did an additional cleaning of the lines based on the characteristics of our particular spectra and the performance of our method, rejecting lines which systematically gave non-satisfactory fits. The selected spectral lines and their atomic data are specified in Table~\ref{tab:lines}.

The spectral fitting was done by comparing the observed fluxes, weighted by their uncertainties, with a synthetic spectrum. This fitting was only done inside the linemask regions, which were defined to be $\pm0.2$~nm around the central peak of the selected spectral features. To perform the fits of each star, the atmospheric parameters (\teff, \logg) were fixed to the ones obtained in Section~\ref{sec:results}, the macroturbulent velocity (\vmac) was fixed to 0 \kms\ so that the only broadening parameter considered is \vsini\ \citep{2018PASJ...70...91T,2022arXiv220610986F} to allow some freedom in the broadening description to fit the lines. We checked that the resulting values of  \vsini\ were compatible with those determined with SIR. We also let the parameters [$\alpha$/M] and \vmic\ vary freely for each spectra. Absolute chemical abundances were then obtained for each spectral line by varying the abundance of the synthetic spectrum until convergence was reached using a least-squares algorithm. 

\subsection{Results}\label{sec:results_abund}
The final abundance for each spectrum was computed as the mean of the individual line abundances, and the uncertainty was computed as their standard deviation ($\sigma$). The Solar abundances were taken from \citet{Grevesse+2007} to obtain the [X/H] values\footnote{[X/H] is the abundance with respect to the Sun, defined as $\mathrm{[X/H]} = \log\left(\frac{N_X}{N_H}\right) - \log\left(\frac{N_X}{N_H}\right)_{\odot}$, where $N_X$ and $N_H$ are the number of absorbers of the element atoms, and of H, respectively.}. For stars observed with two or more spectrographs, there is a general agreement in the abundances, and we chose to use the results of the spectrum with the highest SNR. We list in Table~\ref{tab:results} the resulting mean abundances for our sample. 

From the 20 A stars analysed we obtained an overall $\approx$Solar metallicity with a median and median absolute deviation (MAD) values of $0.08$$\pm$$0.10$ dex. Our result is consistent with other estimates of the metallicity of the cluster in the literature, such as that of \citet{2021A&A...656A.149A}, who obtained [Fe/H]~$=-0.07$$\pm$$0.06$ dex from the analysis of FG stars.

\section{Discussion}\label{sec:discussion}
\subsection{Am candidates in Stock 2}
Am stars are usually identified by their chemical characteristics, particularly an underabundance of some light elements (e.g., O, Ca, Sc) and an overabundance of heavy/neutron-capture elements  \citep[e.g., Ba, Y, Sr;][]{1970PASP...82..781C,pleiades,coma,Hyades}. They also tend to have slow rotational velocities compared to normal A stars. We therefore examined the abundances and $\vsini$ measurements of the stars in our sample to determine the strongest Am star candidates.

\begin{figure}[!thp]
\centerline{\includegraphics[trim=.4cm .05cm .1cm .1cm, clip=True, width = 0.99\columnwidth]{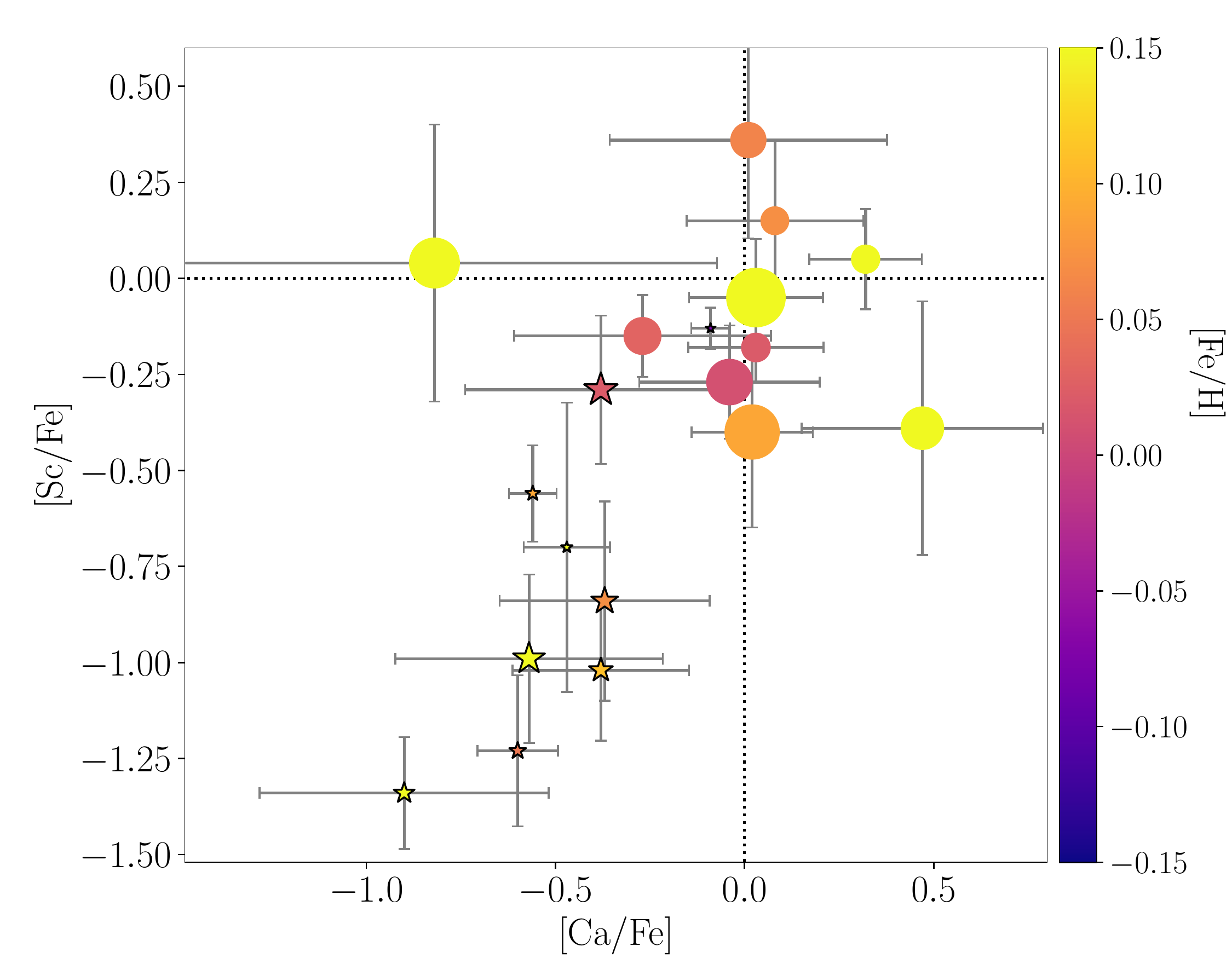}}
\caption{Sc vs.~Ca abundances for our sample, color-coded by the [Fe/H] abundance. The dotted lines indicate values of 0 dex, for reference. The size of the symbols corresponds to rotational velocity for each star; larger symbols are larger \vsini. The star symbols indicated the nine Am candidates listed in Table~\ref{tab:Am}. 
}\label{fig:ScCa}
\end{figure}

We show in Figure~\ref{fig:ScCa} the [Sc/Fe] vs.~[Ca/Fe] abundances, colored by the [Fe/H] values, for our sample. There is a clear correlation between the Sc and the Ca underabundances, with a ``tail" of stars with negative abundances. The star symbols in Figure~\ref{fig:ScCa} are the nine stars that have underabundances of [Sc/Fe] and [Ca/Fe] with a significance of at least 1$\sigma$. Our abundances in these two elements are generally precise compared to other elements, so we use their abundance values as our primary criterion to identify the chemically peculiar stars.

The size of the symbols in Figure~\ref{fig:ScCa} is scaled by the rotational velocity of the stars. The nine stars with negative abundances in [Sc/Fe] and [Ca/Fe] have low rotational velocities: their  $\vsini \lesssim 50$~\kms. Eight of them also present a [Fe/H]~$>0$ dex; the only exception is the star with the smallest underabundance of both Sc and Ca, which has [Fe/H]~$=-0.08\pm$0.05~dex. These stars also have enhanced abundances of neutron-capture elements: all of them have [Ba/Fe]~$>0.3$ dex (reaching values of 0.9 dex), [Y/Fe]~$>0.5$ dex (except the mentioned iron poor star), and [Zr/Fe]~$>0.3$ dex (except one case). 

We consider these nine stars the strongest candidate Am stars in this cluster, and list them in Table~\ref{tab:Am}. Additionally, we identify five stars which have 1$\sigma$ underabundances of Ca or Sc, which we label potential Am stars and also include in Table~\ref{tab:Am} for reference.

\begin{table}[htp]
\centering
\caption{Candidate Stock 2 Am stars and their atmospheric parameters. Above the line are our strongest candidates; below the line are less clear-cut cases. Stars are identified by their BD or Tycho name, and their Gaia DR2 source ID.}\label{tab:Am}
\setlength\tabcolsep{4pt}
\begin{tabular}{lcccccc}
\hline
Name & Source ID & \teff & \logg & \vsini \\
\hline
BD+58 393 & 507310443112918784 & 8631 & 4.05 &	23.5 \\
BD+57 571 & 458857954974047360 & 8072 & 4.28 &	52.3 \\
BD+59 437 & 507315111734402944 & 8768 & 4.00 &	47.9 \\
BD+58 478 & 459067441998085376 & 8871 & 3.97 &	 6.8 \\
BD+59 418 & 507532338299505920 & 9205 & 4.67 &	 8.7 \\
BD+58 423 & 507252993628942720 & 9004 & 3.68 &	36.0 \\
BD+59 428 & 507514058918803328 & 8440 & 3.61 &	16.6 \\
BD+59 438 & 507315219116671488 & 9114 & 3.86 &	13.7 \\
BD+59 431 & 507312539056919296 & 8477 & 3.98 &	30.0 \\
\hline
BD+58 431 & 507270860693863552 & 8907 & 3.35 & 90.2 \\
BD+58 442 & 459223645670638080 & 9290 & 3.64 & 54.0 \\
TYC 3698-2336-1 & 459105306430577408 & 8101 & 3.68 & 77.2 \\
TYC 3698-104-1 & 506844181456308736 & 8424 & 3.94 & 103.9 \\
TYC 3697-1262-1 & 507116585468397056 & 8968 & 3.70 & 35.2 \\
\hline
\end{tabular}
\end{table}

\begin{figure*}[!th]
\centering
\includegraphics[trim=.2cm .3cm 1.cm .8cm, clip=True, width = \textwidth]{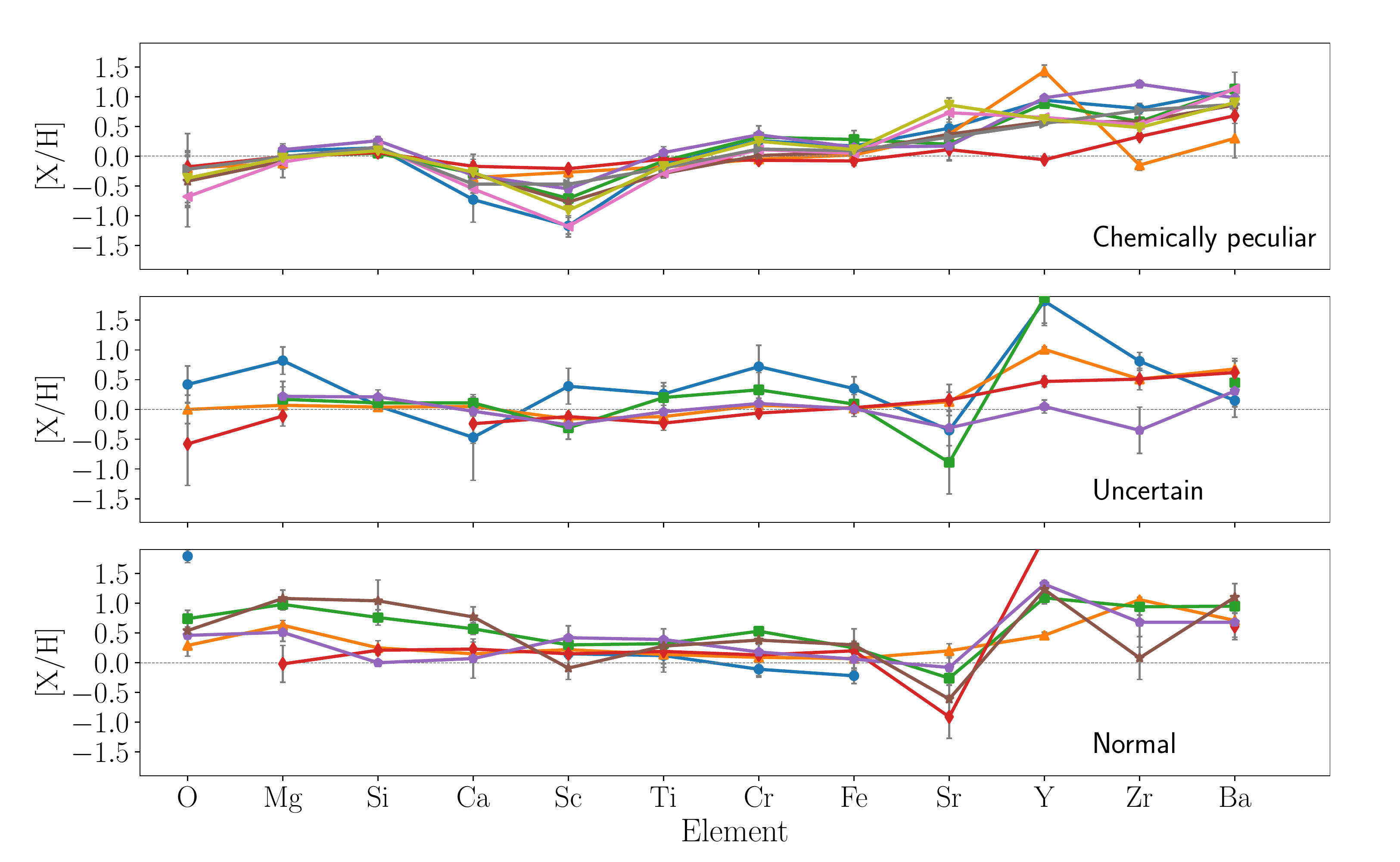}
\caption{Abundances obtained for the full sample of stars that was analysed. The stars are sorted based on the classification described in the text: from top to bottom, Am candidates (note the depletion in Sc, Ca, and O), uncertain cases, and normal stars.}\label{fig:allchemical}
\end{figure*}

In Figure~\ref{fig:allchemical}, we plot the chemical patterns for all the A stars we analyzed.
We distinguish between the  chemically peculiar stars (top panel; nine stars) and normal ($\approx$Solar) stars (bottom panel; six stars), and also include a few cases where the classification is uncertain (middle panel; five stars).

The chemically peculiar stars (top panel) present a noticeable depletion in Sc, Ca, and O:  [Sc/Fe]~$=-0.83$$\pm$0.34 dex, [Ca/Fe]~$=-0.53$$\pm$0.22 dex, and [O/Fe]~$=-0.41$$\pm$0.22 dex. In addition, the abundances we measured for the heaviest elements are significantly enhanced for these stars: [Sr/Fe]~$=0.28$$\pm$0.27 dex, [Y/Fe]~$=0.74$$\pm$0.40 dex, [Zr/Fe]~$=0.40$$\pm$0.35 dex, and [Ba/Fe]~$=0.71$$\pm$0.26 dex. All of the quoted statistics represent the weighted mean and standard deviation of the stellar abundances.

For the stars not classified as Am, 
the abundances are roughly Solar, with a large dispersion between the different stars. We noticed, however, a persistent enhancement of Y, Zr, and Ba similar to what we see in the Am stars: [Y/Fe]~$=0.88$$\pm$0.55 dex, [Zr/Fe]~$=0.49$$\pm$0.35 dex, and [Ba/Fe]~$=0.53$$\pm$0.26 dex. There also appears to be a possible depletion of Sr ([Sr/Fe]~$=-0.32$$\pm$0.45 dex) and a slight enhancement in the abundance of some of the light elements ([O/Fe]~$=0.56$$\pm$0.81 dex and [Mg/Fe]~$=0.31$$\pm$0.35 dex), but these measurements are less statistically significant.

\subsection{Comparison to Am populations in clusters with different ages}
As chemical diffusion is a time-dependent process, models predict that the Am phenomenon should appear once stars are older than about 100 Myr, and that the chemical abundances of Am stars should change as a function of time. Analyzing stars in clusters of different ages helps to examine this hypothesis and can inform the relevant models \citep[e.g.,][]{pleiades,coma,Hyades}. 

\begin{figure*}[htp]
\includegraphics[width = 0.99\textwidth]{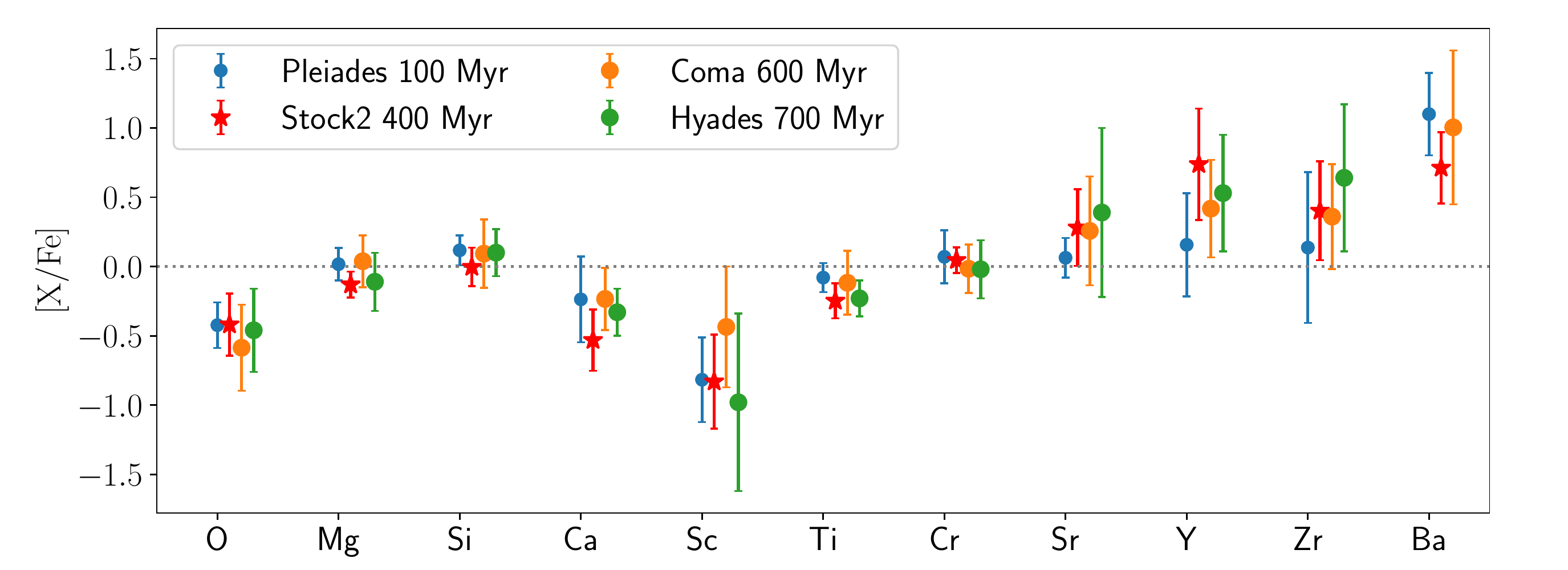}
\caption{Mean abundances of the Am stars identified in the Pleiades, Coma Ber, and Hyades, and the mean abundances of the nine Am stars identified in this work in Stock 2. }\label{fig:clusters}
\end{figure*}

We compared the mean abundances of the Am stars identified in Stock 2 with those of three other clusters: Pleiades, Coma Berenices (Coma Ber), and Hyades \citep[as studied in][respectively] {pleiades, coma, Hyades}. The Pleiades (Melotte 22) is generally thought to be about 100 Myr old \citep[e.g.,][]{Cantat-Gaudin+2020}, and three Am stars were identified by the cluster in \cite{pleiades}. Coma Ber  (Melotte 111) is older, around 600 Myr old, and the Hyades (Melotte 25) is the oldest, at around 700 Myr \citep{douglas2019}. 
Seven Am stars were identified in each of these two clusters. Finally, Stock 2, as discussed above, has an intermediate age of around 400 Myr.

We show in Figure~\ref{fig:clusters} the mean Am [X/Fe] abundances of the four clusters for 11 elements: O, Mg, Si, Ca, Sc, Ti, Cr, Sr, Y, Zr, and Ba.\footnote{We lack Ba measurements for the Hyades.} The error bars represent the standard deviation of the abundances among the different Am stars identified. We see a remarkable level of consistency in the abundance values between the four clusters, and therefore no obvious dependence of the abundances on age.

In particular, we see very similar levels of depletion in O, Ca and Sc, and of  enhancement in the neutron-capture elements Sr, Y, Zr, and Ba, in Am stars in the four clusters. For the heavy elements, there also seems to be an increasing enhancement with increasing atomic number. Am stars in the Pleiades, the youngest cluster, appear to have slightly lower levels of enhancement for Sr, Y, and Zr relative to the Am stars in the three other clusters. However, the uncertainties on the measurements are large, and no clear conclusions with regards to a dependence on age can be made.

Still, the fact that we do not see an obvious age dependence in the depletion and enhancement of the different elements considered across the four clusters is remarkable. In addition to the measurement uncertainties, a number of other factors may explain this seeming inconsistency with theoretical expectations.

Although the slow rotation of Am stars facilitates the occurrence of diﬀusion, other scenarios tried to explain the cause of the peculiarities. \cite{2006PASP..118..419B} suggested that the overabundances of heavy elements observed for the photospheres of Am stars are mainly due to accretion of interstellar gas, dust, and grains, but with very small amounts of hydrogen, which is mainly propelled away by an unstructured stellar magnetic field.
They also added that the low abundances of elements with ionization energies close to those of hydrogen or helium can be explained by charge-exchange reactions with high-energy protons or helium ions that were accelerated by the magnetic fields. Also, \citep{1989Ap.....31..725L} suggested that the orbital motion in the short periodic binary can explain the observed variation of average abundance with time. From the theoretical point of view, most authors agree on the fact that we should consider the simultaneous action of diffusion, mixing, accretion, charge-exchange reactions, winds, mass loss and magnetic field effects.

From the observational point of view one has to consider that the fundamental parameters of the Am stars in Pleiades, Coma Ber, and the Hyades were derived using the UVBYBETA calibration of Str\"{o}mgren photomery \citep{1993A&A...268..653N}, and that the abundances were found using the procedure of \citet{1995PASJ...47..287T}. In contrast, the analysis of Stock 2 we performed relied on SIR for the atmospheric parameters and SPECTRUM (through \ispec) for the abundances. These differences might be enough to bias the comparison presented in Figure~\ref{fig:clusters}, and could hide a small time dependence (regardless of measurement uncertainties).

Additionally, the age range for which we have data remains limited, and a larger sample of clusters is necessary to truly examine the effects of the diffusion mechanism on the studied elements between 100 Myr and 1 Gyr. Accurate membership catalogs with high-quality photometry like those based on Gaia data allow for more precise age determinations, and thus offer the possibility to select the most appropriate clusters to study the Am phenomenon. Complementary high resolution spectroscopic follow-up of high-confidence members is essential to get detailed abundances, and opens the door to revisiting the impact of stellar evolution on the properties of intermediate-mass stars across a wide range of ages and masses, and to comparing observations with the predictions of theoretical models.

\section{Conclusion}\label{sec:conclusion}
The identification of chemically peculiar stars in open clusters presents an opportunity to study the effect of diffusion mechanisms in the atmospheres of intermediate-mass MS stars. Stock 2 is the perfect target to search for chemically peculiar stars, since it is a relatively rich ($>$1000 members) and nearby ($\approx$400 pc) cluster with an age between that of benchmark clusters such as the Pleiades and Hyades.

We performed a spectroscopic study of 71 upper MS members of the cluster. Sixty-four of these stars have spectra from our own observations with the ESPaDOnS and SOPHIE spectrographs, and 13 from archival data taken with HARPS. We derived the fundamental atmospheric parameters \teff, \logg, \met, and \vsini\ for these stars using a combination of  PCA and SIR.

We then obtained 1D-LTE chemical abundances of 12 chemical species (O, Mg, Si, Ca, Sc, Ti, Cr, Fe, Sr, Y, Zr, and Ba) using spectral synthesis. Abundances derived for 20 stars in our sample show that the cluster has an overall $\approx$Solar metallicity, consistent with what is derived in the literature \cite[e.g.,][]{2021AJ....161....8Y, 2021A&A...656A.149A}, but we find a large scatter in Fe abundances.

We identified nine candidate Am stars in Stock 2 by examining patterns in the abundances we measure. These stars present a significant depletion in both Sc and Ca, and an overabundance in the heavy elements Ba, Y, and Zr. These stars also have relatively low rotational velocities, with \vsini\ $\lesssim 50$ \kms. We identified five more stars whose spectra suggest they are underabundant in Ca or Sc, but whose classification as Am stars is less certain. 

The Am stars in Stock 2 present very similar abundance patterns to the Am stars identified in the literature in the Pleiades, Coma Ber, and Hyades open clusters. For example, Am stars in all four clusters appear to have enhancements of neutron-capture elements that increase with increasing atomic number. Furthermore, our data do not allow us to identify a clear time-dependence in the process responsible for the relative depletion or enhancement of a given element---the abundances for all 11 elements we compare are very consistent from cluster to cluster. Our data are insufficient to draw strong conclusions from this observation, but motivate our desire to expand the sample of well-characterized Am stars in which to examine abundance evolution. Having high resolution spectra of A stars members of open clusters well separated in age will allow us to constrain the time evolution of the chemical peculiarities and better understand the source of the Am phenomenon and the interplay between the different mechanisms (diffusion, mixing, accretion, among others). 

Fortunately, Gaia's recent data releases have provided us with an extensive set of new, or re-discovered, nearby and co-eval stellar populations with which to test many dimensions of stellar evolution across a wide range of masses and ages. Our understanding of the physics driving the Am phenomenon is sure to benefit from follow-up campaigns targeting these groups.

\section*{Acknowledgements}
We thank the anonymous referee for the useful comments and suggestions. We thank Javier Alonso-Santiago for kindly sharing the HARPS spectra with us.

This work has made use of data from the European Space Agency (ESA) mission Gaia (\url{http://www.cosmos.esa.int/gaia}), processed by the Gaia Data Processing and Analysis Consortium (DPAC, \url{http://www.cosmos.esa.int/web/gaia/dpac/consortium}). We acknowledge the Gaia Project Scientist Support Team and the Gaia DPAC. Funding for the DPAC has been provided by national institutions, in particular, the institutions participating in the Gaia Multilateral Agreement.

This research made extensive use of the SIMBAD database, and the VizieR catalogue access tool operated at the CDS, Strasbourg, France, and of NASA Astrophysics Data System Bibliographic Services.

We acknowledge support from ``programme national de physique stellaire" (PNPS) and the ``programme national cosmologie et galaxies" (PNCG) of CNRS/INSU. M.A.A.~acknowledges support from a Fulbright U.S.~Scholar grant co-funded by the Nouvelle-Aquitaine Regional Council and the Franco-American Fulbright Commission.

Based in part on observations made at Mauna Kea Observatory (USA), and Observatoire de Haute Provence (France).

The authors wish to recognize and acknowledge the very significant cultural role and reverence that the summit of Maunakea has always had within the indigenous Hawaiian community.  We are most fortunate to have the opportunity to conduct observations from this mountain.


\bibliography{references_1,biblio2_v2}

\end{document}